\documentclass[fleqn,twoside]{article}
\usepackage{espcrc2}

\usepackage{graphicx}
\usepackage[figuresright]{rotating}

\newcommand{\AmS}{{\protect\the\textfont2
  A\kern-.1667em\lower.5ex\hbox{M}\kern-.125emS}}
\newcommand{\ks}{$K_S\rightarrow\pi^+\pi^-e^+e^-~$}
\newcommand{\kl}{$K_L\rightarrow\pi^+\pi^-\pi^0_D~$}
\newcommand{\kss}{K_S\rightarrow\pi^+\pi^-e^+e^-}
\newcommand{\kll}{K_L\rightarrow\pi^+\pi^-\pi^0_D}

\newcommand{\piee}{$K^\pm\rightarrow\pi^\pm e^+e^-~$}
\newcommand{\pimm}{$K^\pm\rightarrow\pi^\pm \mu^+\mu^-~$}
\newcommand{\pill}{$K^\pm\rightarrow\pi^\pm \ell^+\ell^-~$}
\title{CPV tests with rare kaon decays }

\author{E. Marinova\address{INFN Sezione di Perugia, Via A. Pascoli, Perugia, Italy} on behalf of the NA48/1 and NA48/2 collaborations}
       
\begin{document}

\begin{abstract}

The \ks decay mode has been investigated using the data
collected in 2002 by the NA48/1 collaboration. With about 23k signal
events and  59k \kl normalization decays, the
\ks branching ratio was determined.
This result is also used to set an upper limit on the presence
of E1 direct emission in the decay amplitude. The CP-violating
asymmetry has been also measured.

We report on measurements of the rare decays \piee and
\pimm.  The full NA48/2 data set was analyzed, leading
to more than 7200 reconstructed events in the electronic and more than
3000 events in the muonic channel, the latter exceeding the total  existing
 statistics by a factor of four. For both channels the selected 
events are almost
background-free. From these events, we have determined the branching
 fraction and form factors of \piee using different 
 theoretical models.
Our results improve the existing world averages significantly. In 
addition,
we measured the CP violating asymmetry between $K^+$ and $K^-$ in this 
 channel
to be less than a few percent.
\vspace{1pc}
\end{abstract}

\maketitle

The NA48 experiments have a long and successful history in studying direct CP violation effects in the kaon system. The NA48 experiment started collecting $K_L$ and $K_S$ decays in 1997 in order to measure $\epsilon'/\epsilon$~\cite{na48}. In 2002, the experiment continued with a high intensity $K_S$ program aiming to measure the rare $K_S$ and hyperon decays (NA48/1). In 2003, a new beam line delivering simultaneously $K^+$ and $K^-$ was introduced with the goal of measuring the charge asymmetry in $K^\pm\rightarrow3\pi$ decays~\cite{k3pi}. NA48 is a fixed target experiment at SPS -CERN. The main components are a magnetic spectrometer to measure charged particle momenta and an electromagnetic calorimeter based on liquid krypton for measurement of the electomagnetic showers. More detailed information about the detector componets and performance can be found elsewhere~\cite{na48det}.

Studies of the \ks decay have recently been completed. This decay provides a testing ground for a CP non-invariance. The decay amplitude of \ks  is expected to be dominated by the CP-even inner
bremsstrahlung transition~\cite{kstheory}. 
As $K_S$ are mostly CP - even, no contribution from a CP-odd direct emission is expected. Therefore, 
the CP violating asymmetry, defined as \\$A_{\phi} = \frac{ N_{\pi\pi ee}(\sin\phi\cos\phi > 0) - N_{\pi\pi ee}(\sin\phi\cos\phi < 0)}{ N_{\pi\pi ee}(\sin\phi\cos\phi > 0) + N_{\pi\pi ee}(\sin\phi\cos\phi < 0)}$, where $\phi$ is the angle between the $\pi^+\pi^-$ and the $e^+e^-$ decay planes in the kaon centre of mass, is expected to be 0.

The first observation of \ks was by the NA48 experiment in 1998 based on 56 events. From the full 1998 - 1999 data set, a total amount of 677 events was collected. The branching ratio was measured to be BR (\ks) = $(4.69\pm0.30)\times10^{-5}$~\cite{ks99}, and the CP violating asymmetry, $A_\phi = (-1.1\pm4.1)\%$~\cite{ks99}, was found to be compatible with 0. Using the 2002 sample, we acquired more than 20 000 events. A possible contribution from E1 direct emission was investigated. The \ks BR was measured with respect to the \kl channel.

A Geant3~\cite{geant} based Monte Carlo simulation was used for acceptance calculations and most of the background estimation. 
The PHOTOS code~\cite{photos} was implemented in the simulation program to take into
account radiative effects in the acceptance calculation for both signal and normalization
channels.
The simulation includes Coulomb corrections as well. The beam 
           shape was tuned with $K_L\rightarrow \pi^+\pi^-\pi^0$ decays.
In total, 22966 \ks candidates, with a background of 103 events, most of which come from the normalization channel, were collected. The selection of \kl
events is very similar to the one for the \ks. We required, in addition to the four identified charged particles, the presence of a well defined photon giving an in-time signal in the LKr calorimeter, with a minimum energy of 2 GeV. 58983 \kl events were reconstructed and the background contamination was
estimated to be smaller than 0.1\%.

The measured ratio is $\frac {BR (\kss)}{BR(\kll)} = \frac{N(\kss)A(\kll)} {N(\kll)A(\kss)}  R_{\epsilon} R_K$, where N(\kl), N(\ks) is the number of events after background subtraction; A(\kl), A(\ks) are the acceptances, $R_\epsilon $ is the trigger efficiency ratio, and $R_K$ is the $K_L/K_S$ flux ratio. The average acceptances for \ks and \kl are (2.804 ± 0.006)\% and (1.644 ± 0.002)\%. The average value of $R_K$, over the investigated energy range, was
computed to be 0.142 and the average value for $R_\epsilon = 1.023\pm0.018$.

The analysis is performed in 10 bins of the energy of the kaon from 60 to 160 GeV in order to avoid any possible bias due to dependency on the kaon energy spectrum. The result is obtained from fitting the data with a constant parameter, BR (\ks)/ BR(\kl)
= $(3.28 \pm 0.06_{stat}+0.04_{syst}) \times 10^{ -2}$ with $\chi^2~$/~ndf = 8.8~/~9, assuming no contribution from E1 direct emission.The statistical error on the ratio is dominated by the uncertainty on the trigger
efficiency. The main systematic sources are listed in Tab.~\ref{tab:2} with the dominating contributions originating from the 
geometrical cuts and rejection of 
pion decays due to inefficiencies  
in the muon detector and to the fact that
pion decays occurring downstream of the magnetic spectrometer were not included in the
simulation.
\begin{table}
  \caption{Systematic uncertainties to BR(\ks)/BR(\kl)}
 \label{tab:2}
  \begin{tabular}{l l }
Source & $\sigma$ syst
(\%)\\
\hline
$K_L\rightarrow\pi^+\pi^-\pi^0$ matrix element & $\pm$ 0.2\\
Background subtraction & $\pm$ 0.1\\
Radiative corrections & $\pm$ 0.4\\
Trigger efficiency & $\pm$ 0.4\\
e -- $\pi$ separation & $\pm$ 0.2\\
$\pi$ decay & $\pm$ 0.6\\
Beam parameters & $\pm$ 0.1\\
Geometrical cuts & $\pm$ 0.7\\
$K_{L,S}$ lifetimes & $\pm$ 0.3\\
Kinematical cuts & $\pm$ 0.3\\
Reconstruction & $\pm$ 0.3\\
\hline
Total & $\pm$ 1.2
  \end{tabular}
\end{table}

\begin{figure}[htb]
\vspace{9pt}
\includegraphics[width=18pc]{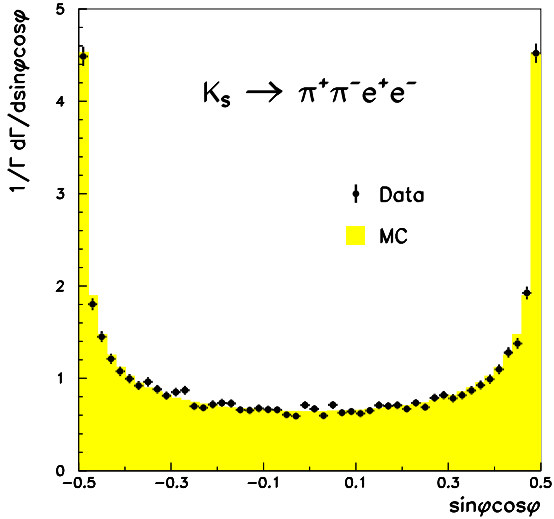}
\caption{The \ks normalized differential decay rate after background subtraction and acceptance correction.}
\label{fig:1}
\end{figure}

Using the PDG value ~\cite{pdg} for the BR (\kl), the following result is obtained BR(\ks) = $(4.93 \pm 0.14) \times10 ^{-5}$. Both, BR (\ks)/BR(\kl) and BR (\ks) results are in agreement with the previously measured results by the NA48 experiment~\cite{ks99}.

Using the relation BR$(K_L\rightarrow \pi^+\pi^-e^+e^- )/ BR(\kss ) = |\eta_{+-}|2(\tau_L/\tau_S)$,
the CP-violating inner bremsstrahlung part of
the analogous $K_L\rightarrow \pi^+\pi^-e^+e^-$ decay is determined to be
BR $(K_L\rightarrow \pi^+\pi^-e^+e^-) = (1.41 \pm 0.04) \times 10^{-7} $, compatible with the theoretical predictions of Sehgal and Wanninger~\cite{sw}.

Following the formalism in~\cite{sw}, a direct emission term was added to the matrix element of the decay in the simulation, $M = e \{g_{BR}e^{i\delta_0} [p_{+\mu} / p_+ \cdot k - p_{-\mu} / p_- \cdot k] + g_{E1}e^{i\delta_1} p_- ( \cdot k)p_{+\mu} - p_+ ( \cdot k)p_{-\mu} \}\{ u (k^-)\gamma ^\mu\nu (k^+) /k^2\}$,
where $e$ is the electric charge, $p_+, p_-, k_+, k_-$ are the 4-momenta of the $\pi^+, \pi^-, e^+, e^-$
particles, respectively. $\delta_0$ is the $\pi\pi$ scattering phase in the I = J = 0 state and $\delta_1$ is the $\pi\pi$ scattering phase in the I = 1 p-wave state.
 $g_{E1}$ is the parameter that gives the magnitude of E1 direct emission and
the $g_{BR}$ parameter is related to the $K_S\rightarrow \pi^+\pi^-$ decay width by $\Gamma(K_S\rightarrow \pi^+\pi^-) = \frac{g^2_{BR}}{16\pi M_K}\left[1-\frac{4M^2_\pi}{M^2_K}\right]^{1/2}$.
To extract the $g_{E1}$ parameter, a fit to the
background-subtracted energy spectrum of the emitted virtual photon in
$K_S\rightarrow\pi^+\pi^-\gamma^*$ was performed
by varying in the Monte Carlo simulation the contribution of $g_{E1}$ with
respect to $g_{BR}$. The best-fit estimation of $g_{E1}/g_{BR}$ is: $g_{E1}/g_{BR} = 1.5 \pm 1.1 $
with a $\chi^2/ndf$ value of 12.8/17. This result is consistent with no observation of E1 direct
emission in the \ks decay.

Finally, the CP-violating asymmetry was measured. The normalized differential decay rate dependence on $\sin \phi \cos\phi$ is Shown in Fig.~\ref{fig:1}  The corresponding
asymmetry parameter was found to be consistent with zero:
$A_{\phi} = (-0.4 \pm 0.7_{stat} \pm 0.4_{syst})\%$.

\begin{table*}
  \caption{Results of fits to the four models and the BR of $K^{\pm}\rightarrow \pi^{\pm}\ell^+\ell^-$ decays.}
\newcommand{\m}{\hphantom{$-$}}
\newcommand{\cc}[1]{\multicolumn{1}{c}{#1}}
 \label{tab:1}
\begin{tabular}{@{} llllll}
 \hline
Model & Parameter & Results & $\chi^2/ndf$  &  Results  &  $\chi^2/ndf$ \\
&  & $K^{\pm}\rightarrow \pi^{\pm}e^+e^-$ &of the fit& $ K^{\pm}\rightarrow \pi^{\pm}\mu^+\mu^-$ &of the fit\\
\hline
 &$ \lambda$ &  2.32 $\pm$ 0.18&& 3.11 $\pm$ 0.56&\\
Model 1  & $|f_0|$ &  $0.531 \pm 0.016$ &  22.7 / 19.0 &0.470 $\pm$ 0.039&12.0 / 15.0\\
\hline
 & $a_+$ & -0.578 $\pm$ 0.016& &-0.575 $\pm$ ±0.038&\\
Model 2  & $b_+$ &  -0.779 $\pm$ 0.066& 32.1 / 19.0&-0.813 $\pm$ 0.142& 14.8 / 15.0\\
\hline
&$\tilde{w}$& 0.057 $\pm$ 0.007 &&0.064 $\pm$ 0.014&\\
Model 3 &$\beta$&0.531 $\pm$ 0.016& 27.7 / 19.0&0.064 $\pm$ 0.014& 13.7 / 15.0 \\
\hline
&$M_a$&0.974 $\pm$ 0.035&&1.014 $\pm$ 0.090&\\\
 Model 4 &$M_\rho$ & 0.716 $\pm$ 0.014& 36.9 / 19.0& 0.725 $\pm$ 0.028& 15.4 / 15.0\\
\hline
Combined result &BR  & $(3.11 \pm 0.12) \times 10^{-7}$&&--&\\
\hline
Model independent &BRmi & $z > 0.08$&& full range&\\
 & & (2.28 $\pm$ 0.08) $\times 10^{-7}$ && $(9.25 \pm 0.62)\times10^{-8}$&\\
\hline
  \end{tabular}
\end{table*}

Other new CPV results from rare kaons come from the \pill decays, measured by NA48/2. The differential rate for these decays depends on the form factors, for which there are 4 models: linear, CHPT at NLO~\cite{chpt98}, ChPT-Large-Nc QCD Model~\cite{friot}, Mesonic CHPT~\cite{pervushin}. Each of the models has two free parameters which can determine a model dependent BR ratio. 

The measurement of the  $K^{\pm}\rightarrow \pi^{\pm}e^+e^-$ decay is based on 7253 events, with a background of $(1.0 \pm 0.1)\%$. The very similar decay $K^\pm\rightarrow\pi^\pm\pi^0_D$, where $\pi^0\rightarrow e^+e^-\gamma$, was chosen as a normalization channel.
The accessible kinematical region in $z$ is above $z<0.08$ due to the presence of background coming from the normalization channel which cannot be efficiently suppressed. The reconstructed $d\Gamma_{K_{\pi ee}}/dz$ spectrum was fitted to the four models, and the form factor parameters were extracted. The four models cannot be distinguished in the visible kinematical region for $K^{\pm}\rightarrow \pi^{\pm}e^+e^-$. However, below $z<0.08$, the theory predicts different behavior of the four models. The form factor fits to the $d\Gamma_{K_{\pi ee}}/dz$ spectrum are reported in Tab.~\ref{tab:1}, together with the model independent BR in the visible kinematic region, and the combined result of the four models for the BR over the whole $z$ range. The results of the first three models and the BR are in agreement with the results reported in~\cite{prevexp1}, \cite{prevexp2},\cite{prevexp3}, and with the theoretical prediction for $a_+= -0.6^{+0.3}_ {-0.6}$~\cite{prades}.  Model 4 was never tested before.

The first measurement of the CP violating asymmetry, done by NA48/2, \\$\Delta(K^\pm_{\pi^{\pm}e^+e^-}) = (-2.2\pm1.5_{stat}\pm0.6_{syst})\times10^{-2}$ is consistent with no CP violation. However, its precision is far from the SM expectation~\cite{cpvpiee}.

The \pimm analysis is based on 3120 reconstructed events, 4 times more than the total world's sample, with a background of $(3.3\pm0.5)\%$. The main technique of background estimation is based on choosing events with two $\mu$ with the same sign from the data sample, and the result is confirmed by a $K_{3\pi}$ MC simulation. 
 For this analysis, the full kinematical region in $z$ is accessible.
Each of the four models for the form factors provides a reasonable fit to the data. The results of the fits are reported in Tab.~\ref{tab:1}. The data sample size is insufficient to distinguish between the models considered.
A measurement of the CP violating asymmetry, $\Delta(K^{\pm}_{\pi^{\pm}\mu^+\mu^-}) = (1.1\pm2.3)\times10^{-2}$, is consistent with CP conservation, but its precision is far from the theoretical predictions~\cite{cpvpiee}.
Another interesting observable, the forward-backward asymmetry in terms of the
$\Theta_{K\mu}$ angle between three-momenta of the kaon and the muon of opposite sign in the $\mu^+\mu^-$
rest frame, was measured for the first time:
$A_{FB} = \frac{(N(\Theta_{K\mu}>0)-N(\Theta_{K\mu}<0))}{(N(\Theta_{K\mu}>0)+N(\Theta_{K\mu}<0))} = (-2.4\pm1.8)\times10^{-2}$, where the error is dominated by the statistical uncertainty.
The achieved precision does not reach the upper limits of the SM~\cite{afbsm} and the MSSM~\cite{afbmssm}, both at the order of $10^{-3}$.
The results on the BR agrees with two of the previous measurements~\cite{pmm:prevexp2}, \cite{pmm:prevexp3}, and disagrees with~\cite{pmm:prevexp1} .
The measurements on the form factors agree with the $K^{\pm}\rightarrow \pi^{\pm}e^+e^-$ results of NA48/2~\cite{piee}, with the $\lambda$ value measured by~\cite{pmm:prevexp2},  and with
theoretical expectation of $a_+= -0.6^{+0.3}_ {-0.6}$~\cite{prades}.

\end{document}